# ACCURATE LOCATION ESTIMATION OF MOVING OBJECT WITH ENERGY CONSTRAINT & ADAPTIVE UPDATE ALGORITHMS TO SAVE DATA

Vijay Bhaskar Semwal[1], K Susheel Kumar[2], Vinay S Bhaskar[3] & Meenakshi Sati[4]

[1]PES-KM Group, Newgen Software Technology, Sector 144, Noida, India
`vijay.semwal@newgen.co.in`
[2]Department of Computer Science & Engg, Ideal Institute of Technology, Ghaziabad
`Sus.iiita.932@gmail.com`
[3]Department of HR & PR ,Electro Equipments, Roorkee.
`vinaysbhaskar@gmail.com`
[4]Department of Computer Science & Engg., Hermes College of Engineering, Roorkee.
`minaxisati@gmail.com`

## ABSTRACT

*In research paper "Accurate estimation of the target location of object with energy constraint & Adaptive Update Algorithms to Save Data" one of the central issues in sensor networks is track the location, of moving object which have overhead of saving data, an accurate estimation of the target location of object with energy constraint .We do not have any mechanism which control and maintain data .The wireless communication bandwidth is also very limited. Some field which is using this technique are flood and typhoon detection, forest fire detection, temperature and humidity and ones we have these information use these information back to a central air conditioning and ventilation system.*

*In this research paper, we propose protocol based on the prediction and adaptive based algorithm which is using less sensor node reduced by an accurate estimation of the target location. we are using minimum three sensor node to get the accurate position .We can extend it upto four or five to find more accurate location but we have energy constraint so we are using three with accurate estimation of location help us to reduce sensor node..We show that our tracking method performs well in terms of energy saving regardless of mobility pattern of the mobile target .We extends the life time of network with less sensor node. Once a new object is detected, a mobile agent will be initiated to track the roaming path of the object. The agent is mobile since it will choose the sensor closest to the object to stay. The agent may invite some nearby slave sensors to cooperatively position the object and inhibit other irrelevant (i.e., farther) sensors from tracking the object. As a result, the communication and sensing overheads are greatly reduced.*

## 1. INTRODUCTION

It is always difficult to track a exact position of moving object by sensor network. In this I used protocol to find the exact position based on data aggregation algorithm and voronoi graph[13] .Each Wireless Sensor node has ability of collecting information, processing then, and storing information, and communicating with its nearest nodes .Take more sensor node in group is used to collect information is waste of resource and we have energy constraint always once we deploy sensor in hostile environment .We can use this technique for detection of many disaster and by collecting information we can take decision accordingly. If object is moving it is always costly to collect or maintain data. In order to maintain data on particular node it might possible node can fail in this case we can lose data. The sensors are used to collect information about mobile target position and to monitor their behaviour pattern in sensor field. Intruder surveillance in military regions. Wild animal habit monitoring is emerging as applications of moving targets are. We have following problem to track the moving object. It is a challenging task with below mention points-target detection, localization, data gathering, and prediction. If we consider the above problem in localization we have main problem when we have to focus on only some object and we are using whole network and suppose condition can arise we need to monitor object to long so sensor will always have to wake up to detect a mobile target. It is one of the wastes of costly resource. As energy is main requirement of sensor once it is deployed in field it is not possible to us change batter, therefore each sensor must minimize its battery power usage for desired longevity of network operation, which can be accomplished by properly managing sensor's operation. We have many protocols to monitor moving target [11, 12].Some prediction based protocol [8] we cannot perform well once prediction wrong. [5] Which is based on liner target trejotry. The focus of this paper is to develop a solution which can efficient track the path with minze the participating nodes, we use protocol which are based on prediction of moving object in 3-dimension wireless sensor with linear estimation.

## 2. LITERATURE SURVEY

### 2.1 Currently Existing Technologies-

As we all knows that wireless sensor network is either partially or fully developed and lot of research is still going .As lot of research is going for good MAC layer protocol [2].
**Physical Layer** – Which is basically used to transfer of data and all the physical media come under it. In this layer really data transfer happened.
**Mac Layer**-In this layer we send the data as the form of Frame. It controls the frame and guarantee time slot .So it is guarantee for secure service.

### 2.1.1 Problem definition and scope –

As we know sensor node are using battery power and they communicating with other mote .All the sensor node finally send the data to base node which finally collect on sink .The position of sensor node are unknown. Some application are-

1. Finding the position of moving entity to monitor the animal
2. To control the traffic by monitoring the speedy vehicle.
3. In Building. to find the any fracture and crack.

### 2.1.2 Formulation of the present problem –

When the moving entity is moving and reached in area where sensing range of particular sensor is less that time we need some other sensor node which can take position of that particular sensor but during this many other sensor can also sense the that particular entity to overcome we send a special message which indicate no need to you take place by message passing. But we have to consider some fact like as sensor node ahs energy or sensing restriction so w need to do some trade of between both.

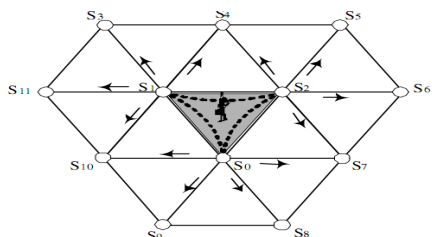

Figure 1. Multiple Sensing Objects

International Journal of Distributed and Parallel Systems (IJDPS) Vol.2, No.4, July 2011

## 2.2 The already existing protocol for location tracking are based on master slave communication in which location tracking

By each sensor node is taking part and working together .When and entity detect in particular range of sensor it start his own SELECTION PHASE[6] process start by one sensor node to find on which simulation engine can be install to track the position. As the entity will visit on another region, the slave point will be called and assign then again responsibility. And here we will take one more point in consideration signal strength thresholds which is used for the purpose when to called and assign back again the responsibility to slave point, any slave point node who did slave job is marked by black. We are using two slave point node but for more accurate location we can use more then 2 sensor node as a slave. To reduce the volume of position data to be propagate on, a master point forward some tracking information to position server.

### 2.2.1 Protocol Details.

In already existing approach sensor node is able to make difference in one entity to another .This is possible by when sensor node will send unique ID code periodically. Even taking multiple entities in consideration but doing processing separately. Here a start transition diagram of sensor node.All sensor node stays in IDLE state in starting and performing protocol. In this state sensor always monitor the entity and one any entity come in the area of sensor it start ELECTION phase for serving as master .The sensor which is nearest one win the process and become the master point and select the two slave point on other two sensor node . Once we have master point it will become master and start working on protocol and same slave will go in slave state and start slave protocol .One the entity enter in the backup area new process start to select master and slave point.

## 3. Propose work

### 3.1 Network Model with 2D plane-

Here we are deploying sensor node in 2D plane i.e. our network structure is 2D plane .We can place sensor in regular or irregular network .Here we are assuming the each sensor node is aware of it's own physical location and neighbor location and each node also using piggy backing concept also to send own data to neighbor node so that if any node fails due to some disaster we can recover data from neighbor node cause node is sending data to neighbors as a piggybacking. We also only need the 3 sensor node to detect the location of moving entity .by three sensor node we can get the exact position of particular entity and once sensor node has detect the position sensor node is sending to server by time to time depending based on priority either it is real time or not. We are dividing all the sensing range into 2 area interested area and supportive area which is used during handover.

### 3.1.1 Work done in research paper

 The previous work motivate to carry this work on more elegant way and expand it upto four sensor node .In this thesis we collect the data from various source and simulate it on trimsim which shows the proper result and indicating the success of thesis.   This paper taking 3 sensor nodes to detect the position of entity and drawing in 2 D plane. We have assumed the some of agent which will work one by one. One node will come another will wait if entity will come in the range of that particular range then that particular sensor node will take responsibility. Once one node had took responsibility it will do whole work with collection of data, passing that data to neighbor and in case of node failure it is taking appropriate action like in case of disaster flying of data from one node to another managing memory and in case of flooding of data removing old data. New cloud computing tool window Azure of Microsoft is used for collecting data and store so it is also helping to reduce the overhead of memory requirement and fear of data loss .Even we do not have much capacity we can direct store on Azure as a sink node.  As I collect data on the three sensor node and simulate on TRMSIM and I also placed the four sensors and collect the data and deploy on the TRIMSIM .It is obviously good to have exact position four sensor id better in compare to 3 sensors. I also observe it there is very less data loose in node failure because we are using piggy backing concept and one more point to notice here in 4 sensor nodes due to less collision of data we are getting good and more close look of entity which is moving. And for future reference my suggestion to have such protocol which can trace the path of highly fast object. One of the important drawback of 3 sensor node is if object is in  the sensing range of two sensor node .It is difficult to sensor who will take the control.  And one more improvement is how to distinguish between moving entity. All the experiment I did and analysis it with 3 sensor node or 4 sensor node .And I tried to keep balance between both of it. I have performed simulation on trimsim-regular release which provides a simulation framework for 3 d sensor network. We have implemented our approach and chord selection approach mentioned in and the comparative results are presented in the subsequent subsections.

### 3.2 Active Update Algorithm

Active Update entails a broadcast of a sensor's measurement to all sensors within its transmission range. This process is recursively continued and enables the sensors to progressively and collaboratively share and piggyback measurements amongst themselves. Resulting in the aggregation of global knowledge at every sensor. Thus, when a sensor floats within transmission range of a Pick-up station, it not only reports its measurement but also the measurements of other sensors (piggybacked onto itself) with which it participated in following step of Active Update Algorithm.

### 3.2.1 Assumption

Consider the case where a set of k targets need to be tracked with 3 sensors per target from the resource requirement viewpoint.  They show that the probability that all targets can be assigned 3 unique sensors shows phase transition properties

### 3.2.2 Protocol for Update

 Here each sensor network is broadcasting measurement to the entire sensor network within transmission range. We do this process recursively which allow sensor to share data with agreement .Here we are using the concept of piggybacking i.e. every one sensor node storing its own information and taking information from other. In our assumption each sensor node is storing the following details 1- Sensor ID, 2 - Value, 3- Data Table, and 4- Flag of Change in table

Here In this describing the step of my Algorithm -

Step 1- In this step quick wake up sensor node will try to sense value and will collect-
      Collect Value ();
Step 2- After sensing value sensor node will send sensing data and id-
      Distribute to all (Table of Data);
Step 3- If sensor is sensing first time then flag should be false
      Changed in Data Table value=False;
Step 4-On receiving a Distributed value containing a Table
      For every data in table i.e. sensor Id and collect value
Step 5- If there is no previous sensing value
      Then add Sensor Id and Collect Value in table
Step 6- If any change will happen i.e.
      Changed in Data Table value==true
Step 7- Distribute

      Distribute (table);
Step 8 –At point if a signal is received from the pointing station,
      Transmit (Value of Table) to pointing station and exit.

For irregular network in which we do not have idea how to choose master and two slaves point for given entity. So here I am applying Voronoi graph problem in geometry. We have set of point P in a 2D plane, by vornoi graph we ill divided in sector. The total no of sector are |p| and each segment have point which are close of it. And point in particular sector will serve as master. We can slave 3D problem by divided and conquer approach..And the other point in region will server as slave point. We will do it repeatedly and each time one node will consume by P-k.

## 4. Simulation

As we are developing our network model on randomly deploy sensor node .My algorithm is simulating on a chunk of randomly generated network. And result is the average. The most attracting feature of my simulation is that, we are assuming sensor node is usually assumed to isotropic. Due to this the range of sensors is perfect circle. Under such an assumption, the Localization results provide the best-case performance for an algorithm.

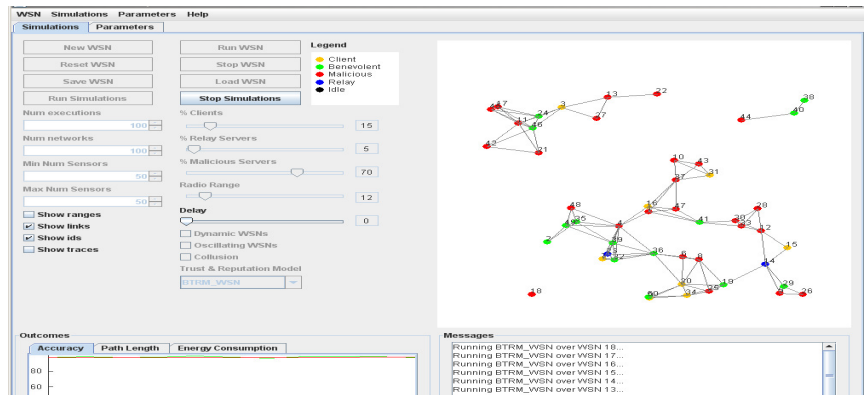

Figure 2. Simulation Result

## 5. Result

As In this performing in 2D plane with 3 sensor node and I collect the data and I found it in 4 sensor node we are getting more accuracy but in 3 sensor node we are getting higher performance cause we are consuming less using less sensor node and finding the position. We deployed whole data gather by me on trimsim which is working on 2 d plane it is basically java applet called in c#.net. We have some more API using and need some jar file to support our application. As I implemented it in 4 layer Business layer, User interface layer, Data Base layer. As In this using 5 parameter taking
Input parameter –

Data In binary Array
Byte [] b=getDataInBinary ();
Location calculate
String strLocation x , strLocation y
And we have two output parameter
Out Graph, param simulation result;
     By using above protocol I find the following desired result. As we have deployed the

thousand of sensor mode randomly and as object is moving these sensors is tracing the path. These graph showing the dotted path of tracing location. We are getting accurate position of object.

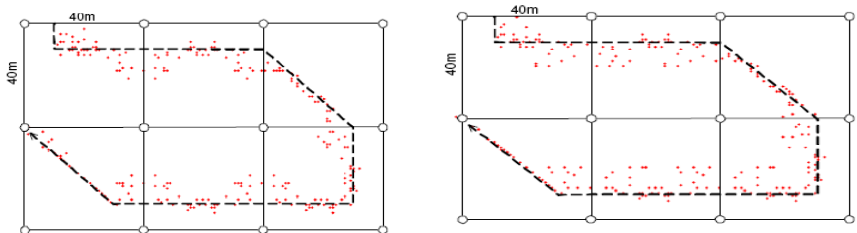

      Figure a.  By 4 sensor node        Figure b By 3 sensor node

Figure 3. Tracking Result

### 5.1 Data Set Using for Simulation result

This simulation result using the following Data Set in which db, tb, tb are the sensor mode with different capability and generating the result on upon data set

Table 1. Data Set of Different Sensor Mote

|   | DB   | TB(t=12) | TB(t=40,000) | NAB |
|---|------|----------|--------------|-----|
| 1 | 5    | 5        | 7            | 8   |
| 2 | 5.67 | 7        | 8            | 15  |
| 3 | 6    | 8.6      | 7.8          | 20  |
| 4 | 6.1  | 9        | 8.2          | 30  |
| 5 | 8    | 10       | 8.4          | 35  |
| 6 | 10   | 12       | 11           | 40  |
| 7 | 11   | 13       | 12           | 55  |

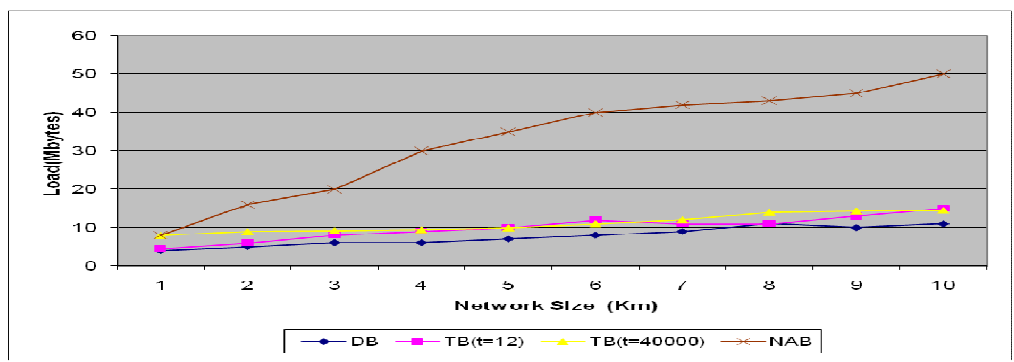

Figure 4. Graph over network size and load of different sensor mote

This graph is indicating the comparison between Load and Network. As the network size of network is increasing load is also increasing .But in this DB is performing well.

## 6. Conclusion

In this Research Paper we propose a method which is using less sensor node reduced by an accurate estimation of the target location. As two sensor node have intersection area which is not giving exact position so we are using minimum three sensor node to get the accurate position .We can extend it up to four or five to find more accurate location but we have energy constraint so we are using three with accurate estimation of location help us to reduce sensor node. We show that our tracking method performs well in terms of energy saving regardless of mobility pattern of the mobile target. Our simulation result shows we are to extend the life time of network with less sensor node and find the exact position of tracking object. It is all regardless of mobility pattern defined by Random Waypoint model and Gauss Markov model [8,9]. Here in this paper I proposed a position-finding protocol for regular or irregular networks. A simulation Point approach is adopted, which enables point to move around to follow the moving agent, so it is greatly reducing the communication and sensing overhead. My protocol is using the approach to send data to next node (near node) so we can save data without losing it. In this using 3 sensor nodes to find the location of moving entity. As we are aware with the cost or other factor of sensor node so in this not using 4 sensor node which is reducing the overhead. As piggybacking is the important concept which I add here by which I not only saved the data of failed node but also reduce the overhead of saving data because we are forwarding the data to neighbour node. So great reduction of overhead of data storage and no overhead of node failure. Once we have successfully started data gathering we can simulate it on simulator then we can find all the desire result.

## ACKNOWLEDGEMENTS

The First I would like to thanks some great mind without whom this research would have been a distant reality. I am totally by the side of these people. I would like to say thanks to my parents who support to me carry out my research without any hindrance. My deepest thanks to great person, my mentor Dr. Shirshu Verma and a big thanks to Mr. K. Susheel Kumar without whose ideas it was impossible last but not least to Mr. Vinay S Bhaskar and to Ms. Meenakshi Sati for excellent analysis of algorithm. I also extend my heartfelt thanks to my well wishers and unmentioned name.

International Journal of Distributed and Parallel Systems (IJDPS) Vol.2, No.4, July 2011

## Authors

**Vijay Bhaskar Semwal** presently working with Newgen Software Technology, Noida, India as a Software Engineer in Research Department. He is M.Tech. From Indian Institute of Information Technology, Allahabad, his major research work interest in Wirless Sensor Network, Artificial Intelligence, Image Processing, Computer Network &Security, Face and Pattern Preconisation and Design & Analysis of Algorithm.

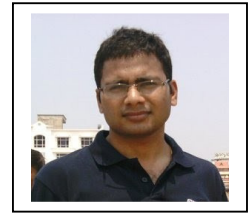

**K Susheel Kumar** presently working as Assistant Professor in Ideal Institute of Technology, Ghaziabad, India. He is M.Tech form Indian Institute of Information Technology, Allahabad, his major research work Interest in Image Processing, computer sensor network & Pattern Recognition

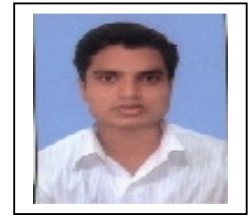

**Vinay S Bhaskar** presently working in Electro Equipment Roorkee(U.A.), India as Astt. Manager (HR&PR). He did PGDMA from I.I.M.S. Pune. His major research area Human and Personal Relationship, and wireless sensor network.

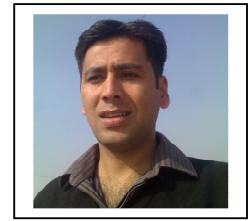

**Meenakshi Sati** presently working as Assistant Professor in Hermas College of Engineering & Management, Roorkee (U.A), India. She is B.Tech from Graphic Era University Dehradun and pursing M.Tech from G.B.Pant Engineering College, Ghurdauri, Pauri Garhwal(UTU). Her major research work Interest in Image Processing, Sensor Network, Design and Analysis of Algorithm, Data Structure.

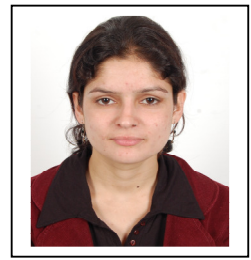